\documentstyle[12pt]{article}

\begin{document}
\newcommand{\gtrsim}{ \mathop{}_{\textstyle \sim}^{\textstyle >} }
\newcommand{\lesssim}{ \mathop{}_{\textstyle \sim}^{\textstyle <} }
\begin{titlepage}
\begin{center}
\hfill    ICRR-Report-345-95-11\\
\hfill    UT-730\\

\vskip 1cm

{\Large \bf Can Decaying Particles Raise the Upperbound on the Peccei-Quinn
Scale?}
\vskip 2cm
{\large M. Kawasaki}\\
{\it Institute for Cosmic Ray Research, University of Tokyo,
  Tanashi 188, Japan}\\
\vskip0.71cm
{\large T. Moroi}\\
{\it Theoretical Physics Group, Lawrence Berkeley Laboratory,
University of California, Berkeley, CA~94720, U.S.A.}
\vskip 0.7cm

{\large T. Yanagida}\\
{\it Department of Physics, University of
  Tokyo, Tokyo 113, Japan}
\end{center}

\vskip 2cm

\begin{abstract}
    We have reexamine the effect of entropy production on the cosmic
    axion density and find that the Peccei-Quinn scale $F_a$ larger
    than about $10^{15}$~GeV is not allowed even if large entropy is
    produced by the decays of coherent oscillations or
    non-relativistic massive particles. We stress that this result is
    independent of the details of models for the decaying
    particles.
\end{abstract}

\end{titlepage}

\section{Introduction}

The axion~\cite{Peccei,Wilczek,Kim,DFS} is the Nambu-Goldstone boson
associated with the Peccei-Quinn symmetry breaking which was invented
as a natural solution to the strong CP problem in
QCD~\cite{tHoot}. The breaking scale $F_a$ of Peccei-Quinn symmetry is
stringently constrained by laboratory experiments, astrophysics and
cosmology. The consideration of cooling processes due to the axion
emission in red giants and SN1987A requires that $F_a$ should be
greater than about $10^{10}$~GeV\cite{Raffelt}.  On the other hand,
$F_a$ should be less than about $10^{12}$GeV so that the energy
density of the coherent oscillations of the axion field should be less
than the critical density of the universe~\cite{Preskill}. Thus the
allowed range of $F_a$ is given between $10^{10}$~GeV and
$10^{12}$~GeV which is called the axion window.

However, Steinhardt and Turner~\cite{steinhardt} showed that the
entropy production due to the first order phase transition or the
out-of-equilibrium decays of massive particles dilutes the axion
density and makes the upperbound on $F_a$ as large as
$10^{18}$GeV. The upperbound ($F_a \lesssim 10^{18}$~GeV) was obtained
by requiring the entropy production should not dilute the baryon
density too much, {\it i.e.} the dilution factor should be less than
$10^6$ assuming the initial baryon-to-entropy ratio to be at most
$10^{-4}$.

However, in the case of the decaying particles, the analysis in
ref.~\cite{steinhardt} is unsatisfactory since the authors assumed the
radiation dominated universe when the axion field starts to oscillate.
As we will show below, it is more reasonable to consider that the
universe is already decaying-particle dominated at that epoch.
Therefore, we reexamine the entropy production from the decays of
coherent oscillations or light particles and obtain the upperbound,
$F_a \lesssim 10^{15}$~GeV, by imposing the reheating temperature to
be less than 1~MeV in order to keep the success for the primordial
nucleosynthesis in the big-bang cosmology. We stress that this result
is independent of detailed models for the decaying particles.

\section{Cosmological Evolution }
\label{sec:evolution}

Let us consider the cosmological evolution of the axion field and the
coherent oscillation of the field $\phi$ with potential
$m_{\phi}^2\phi^2/2$ and density $\rho_{\phi}$. This coherent
oscillation is equivalent to the non-relativistic decaying particle
with the same energy density. Therefore, we only consider the coherent
oscillation hereafter.  We assume that $\rho_{\phi}$ dominates the
energy density of the universe when the oscillation of the axion
starts. If the universe is radiation-dominated, the axion starts to
oscillate at $T\simeq 1$~GeV.  In this case the entropy production
factor is given by $\sim [(T_R^4/T_1^4) (a(T_R)/a(T_1))^4]^{3/4}\simeq
(T_R/T_1)^3 (\rho_{\phi}(T_1)/T_R^4) \lesssim (T_1/T_R)$, where $a$ is
the scale factor of the universe, $T_R$ the temperature just after the
$\phi$ decay and $T_1$ the temperature at which the axion field starts
to oscillate. Since $T_R$ should be greater than 1~MeV to keep the
success of primordial nucleosynthesis, the entropy production factor
becomes less than $O(10^3)$. Therefore, the $\phi$-dominated universe
is a good assumption as far as the large entropy production (with the
entropy production factor greater than $O(10^3)$) is considered.

The axion starts to oscillate at $t=t_1$ when $3H \simeq m_a$. Thus at
$t=t_1$,
\begin{equation}
    \label{start}
    \rho_{\phi}(t_1) \simeq \frac{m_a(T_1)^2M^2}{3},
\end{equation}
where $M=2.4\times 10^{18}$GeV is the gravitational mass and $m_a(T)$
is the axion mass which depends on the temperature $T$
as~\cite{Kolb-Turner}
\begin{equation}
    m_a(T) \simeq
     \left\{ \begin{array}{ll}
    0.1 m_a (\Lambda_{\rm QCD}/T)^{3.7}
    & ~~~{\rm for}~T \gtrsim \Lambda_{\rm QCD}/\pi,\\
    m_a & ~~~{\rm for}~T \lesssim \Lambda_{\rm QCD}/\pi,
    \end{array} \right.
\label{axion-mass}
\end{equation}
where $\Lambda_{\rm QCD}\simeq 0.2$ GeV, and $m_a$ is the axion mass
at $T=0$.

Until the coherent $\phi$ oscillation decays, the ratio of the
axion- to $\phi$-number densities stays constant. Therefore, the ratio
of the energy density of the axion $\rho_{a}$ to that of $\phi$,
$\rho_{\phi}$, is expressed as
\begin{equation}
    \label{axion-phi-ratio}
    \frac{\rho_a}{\rho_{\phi}}
    = \frac{m_a}{m_{\phi}}
    \frac{m_a(T_1)F_a^2\theta^2}{\rho_{\phi}(t_1)/m_{\phi}}
    = \frac{3}{2}\frac{F_a^2\theta^2}{M^2}\frac{1}{\xi(T_1)},
\end{equation}
where $\theta \sim O(1)$\footnote{
$\theta$ takes $\pi/\sqrt{3}$ in the non-inflationary
universe~\cite{Kolb-Turner}.}
is the initial axion amplitude in units of $F_a$ and $\xi(T_1) \equiv
m_a(T_1)/m_a \le 1$.

The $\phi$ decay occurs when $3H \simeq \Gamma_{\phi}$ where
$\Gamma_{\phi}$ is the decay rate of $\phi$. The cosmic temperature
$T_R$ just after the decay is given by
\begin{equation}
    \label{reheating}
    T_R \simeq \left(\frac{10}{\pi^2 g_*}\right)^{1/4}
    \sqrt{M\Gamma_{\phi}}
    = 0.55\sqrt{M\Gamma_{\phi}},
\end{equation}
where $g_*$ is the effective number of massless degrees of freedom and
we have taken $g_* = 10.75$. The entropy density just after the decay
is given by
\begin{equation}
    \label{entropy}
    s(T_R) = \frac{2\pi^2}{45}g_* T_R^3.
\end{equation}
Since the energy density of the coherent $\phi$ oscillation just
before the decay is $M^2\Gamma_{\phi}^2/3$, the axion density at the
decay epoch becomes
\begin{equation}
    \label{axion-density}
    \rho_a(T_R) \simeq \left(\frac{\rho_a}{\rho_{\phi}}\right)
    \rho_{\phi}(T_R) \simeq
    5.3 T_R^4 M^{-2}F_a^2 \theta^2\xi(T_1)^{-1}.
\end{equation}
Then we can estimate the axion-to-entropy ratio as\footnote{
This formula (\ref{axion-entropy-ratio}) is applicable for $T_R
\lesssim 1$~GeV.}
\begin{equation}
    \label{axion-entropy-ratio}
    \frac{\rho_a}{s} \simeq 1.1 T_R F_a^2
    \theta^2 M^{-2}\xi(T_1)^{-1}.
\end{equation}
This value should be compared with the ratio of the present values of
the critical density $\rho_{cr,0}$ to the entropy density $s_0$, which
is given by
\begin{equation}
    \label{critical-entropy-ratio}
    \frac{\rho_{cr,0}}{s_0} \simeq 3.6\times 10^{-9} h^2{\rm GeV},
\end{equation}
where $h$ is the present Hubble constant in units of 100km/sec/Mpc.
Then, the density parameter of axion $\Omega_a$ is expressed as
\begin{eqnarray}
    \Omega_ah^2 &\simeq& 3.1 \times 10^8  {\rm GeV}^{-1}
    T_R F_a^2 \theta^2 M^{-2}\xi(T_1)^{-1}
\nonumber \\
    &\simeq& 5.3 \left(\frac{T_R}{1{\rm MeV}}\right)
    \left(\frac{F_a\theta}{10^{16}{\rm GeV}}\right)^2
    \xi(T_1)^{-1}.
    \label{omega}
\end{eqnarray}

\section{Constraint on $F_a$}
\label{sec:constraint}

Since the entropy production should occur before the primordial
nucleosynthesis, $T_R$ should be higher than about 1 MeV. Then, from
eq.(\ref{reheating}), we get $\Gamma_{\phi} \gtrsim 1.3 \times
10^{-24}$ GeV.  Requiring $\Omega_a h^2 \lesssim 1$ and taking $\xi
\le 1$ into account, we can obtain the upper limit on $F_a$ from
eq.(\ref{omega}):
\begin{equation}
    \label{constraint}
    F_a \lesssim 4.4\times 10^{15} \theta^{-1} {\rm GeV}.
\end{equation}

The above constraint might be more stringent if the cosmic temperature
at $t_1$ is greater than about $0.1$~GeV and hence $\xi \ll 1$.
Therefore, we need to estimate $T_1$ and $\xi(T_1)$.  For this end, we
must take account of the fact that the temperature decreases as $T
\propto a^{-3/8}$.\footnote{
Note that the decay does not occur instantaneously. During the $\phi$
decay, the temperature does not decrease as $a^{-1}$ due to the
heating effect of the decay. For details, see ref.~\cite{Kolb-Turner}.}
Using $a(T_R)/a(T_1) \simeq (\rho_{\phi}(T_R)/\rho_{\phi})^{-1/3}
\simeq (m_a(T_1)/\Gamma_{\phi})^{2/3}$,
\begin{equation}
    \label{T1-TR}
    \frac{T_1}{T_R} \simeq
    \left(\frac{m_a(T_1)}{\Gamma_{\phi}}\right)^{1/4}.
\end{equation}
\hspace{0cm}From eqs.(\ref{axion-mass}), (\ref{reheating}) and
(\ref{T1-TR}), $T_1$ is given by
\begin{equation}
    \label{T1}
    T_1 \simeq 0.07 {\rm GeV} \left(\frac{T_R}{\rm 1MeV}\right)^{0.26}
    \left(\frac{F_a}{10^{15}{\rm GeV}}\right)^{-0.13}.
\end{equation}
When $T_R \simeq 1$~MeV and $F_a \simeq 10^{15}$~GeV, $T_1$ is about
0.1 GeV. At such low temperatures, the axion mass is almost equal to
its zero-temperature value, {\it i.e.} $m_a$. Therefore, $\xi(T_1)
\simeq 1$ and the constraint (\ref{constraint}) is correct.

\section{Conclusion}
\label{sec:conclusion}

We have reexamined the effect of the large entropy production on the
axion density in the early universe and have found that the
Peccei-Quinn scale $F_a$ larger than about $10^{15}$~GeV is not
allowed. This upperbound is three orders of magnitude larger than that
without entropy production but much smaller than the previous
estimation~\cite{steinhardt}. In the present analysis, we have assumed
that the universe is dominated by the coherent oscillation or the
decaying particle when the axion starts to oscillate. On the other
hand the radiation-dominated universe was assumed in the previous
work.  However, when the entropy is increased by a factor greater than
$O(10^3)$, our assumption is correct. On the other hand, in the case
where the entropy production factor is less than $O(10^3)$, the
upperbound on Peccei-Quinn scale $F_a$ is raised up by $O(10^3)$ at
most which completes our conclusion.

We have not discussed the dilution of the cosmological baryon number
asymmetry because it depends on the details of the models for
baryogenesis. For example, if we adopt the Affleck-Dine mechanism for
baryogenesis~\cite{AD}, the $\phi$ decay with the reheating
temperature about 1~MeV is consistent with the observed baryon number
of the universe~\cite{moroi}.

\end{document}